\documentclass[12pt,leqno]{amsart}
\usepackage{graphicx}
\usepackage{indentfirst,csquotes}

\usepackage{amssymb,amsthm,amsmath}
\usepackage{xcolor,paralist,fancyhdr,etoolbox}
\usepackage[numbers]{natbib}
\usepackage{array,tabularx,booktabs,multirow}
\usepackage{caption}
\usepackage{placeins}
\usepackage{hyperref}
\graphicspath{{images/}}

\AtBeginEnvironment{table}{%
  \centering
  \footnotesize
  \setlength{\tabcolsep}{3pt}
  \renewcommand{\arraystretch}{1.15}%
}
\AtBeginEnvironment{table*}{%
  \centering
  \footnotesize
  \setlength{\tabcolsep}{3pt}
  \renewcommand{\arraystretch}{1.15}%
}

\makeatletter
\renewcommand\section{\@startsection{section}{1}{\z@}%
  {2.5ex plus 1ex minus .2ex}
  {1.5ex plus .2ex}
  {\normalfont\bfseries\scshape\raggedright}}%
\renewcommand\subsection{\@startsection{subsection}{2}{\z@}%
  {2.25ex plus 1ex minus .2ex}%
  {1.0ex plus .2ex}
  {\normalfont\bfseries\raggedright}}%
\makeatother

\hypersetup{ colorlinks=false, linkcolor=black, filecolor=black, urlcolor=black }

\usepackage{lipsum}

\begin{document}
\title[Extracting Semantics from Cattle Reporting Categories]{Extracting Semantics from Cattle Reporting Categories for Data Interoperability and Findability} 
\author[Raymond]{Kassy Raymond, Andrew Hamilton-Wright, Deborah Stacey}
\date{\today}
\address{Address}
\email{kassyraymond2@gmail.com}
\maketitle

\let\thefootnote\relax
\footnotetext{} 

\begin{abstract}

Livestock population data disaggregated by age, sex, and production are important inputs to calculations and models that inform our understanding of global health, yet these data are fragmented across disparate sources.  Bridging data siloes to improve the findability of data requires interoperability. Conventional approaches to improving the findability and interoperability of data include indexing standardized metadata. However, creating metadata is time and resource-intensive and is often difficult in domains such as livestock, which lack standards that address the needs of broad user groups. When metadata exist, they typically need to be standardized against a pre-existing vocabulary, ontology, or thesaurus, requiring a technique known as `crosswalking'. To overcome issues in the absence of metadata, the lack of standards, and the resource-intensive solutions that currently exist, this study uses a bottom-up approach. By leveraging real-world reporting categories in datasets, the composition and semantics of terms already present in the data were extracted and analyzed. Using cattle data as a pilot, we find the age, sex, and production modifiers present across cattle terms from five datasets from four data sources capture granularity not present in AGROVOC, the largest agricultural vocabulary in the world. We discuss how the composition and semantics of these terms can be used to improve the interoperability and findability of data without first requiring metadata to be generated or standards to be created. Rather than forcing datasets to conform to an existing vocabulary, this approach uses the semantics embedded in terms already present in datasets, allowing systems to make data more discoverable and interoperable while maintaining culturally and dataset-specific terminology.

\end{abstract}

\section{Introduction}

Climate change, antimicrobial resistance, zoonotic disease, and food security are critical challenges impacting One Health issues spanning human, environmental, and animal health. Understanding how to control, prevent, and mitigate the impacts of these global health crises is paramount to reducing human suffering and improving livelihoods. Intergovernmental bodies, including the Food and Agriculture Organization of the United Nations (FAO), the World Organisation for Animal Health (WOAH), and the World Health Organization (WHO), have recognized that robust computational infrastructure, data governance, and data management are prerequisites for leveraging the datasets needed to understand and monitor these global health issues and to measure progress toward the Sustainable Development Goals (SDGs) \cite{Tripartate2022, FAO2025}. Concurrently, international initiatives such as Livestock Data for Decisions (LD4D) and the Global Burden of Animal Diseases (GBADs) emphasize that interoperability within the livestock sector is critical for accurate epidemiological and economic modelling \cite{LD4D2023, Raymond2023}.

At the national level, livestock data collected through agricultural censuses and surveys provide foundational inputs for modelling greenhouse gas emissions \citep{Herrero2013}, estimating antimicrobial use \citep{Mulchandani2023}, assessing agricultural economics \citep{Huntington2021, Schrobback2023}, and calculating animal production yields \citep{Kebede2024} and biomass \cite{Jemberu2022}. Despite their value, these data remain heavily siloed across disparate repositories and portals, limiting their Findability, Accessibility, Interoperability, and Reusability in accordance with the FAIR Guiding Principles \citep{Cabrera2025, Kebede2024, Wilkinson2016}. These challenges are particularly consequential as the livestock sector increasingly seeks to leverage ``AI-ready'' data through Large Language Models (LLMs) and other advanced AI tools. 

Bridging data siloes so that disaggregated datasets from national and intergovernmental portals can be collectively findable and retrievable requires semantic interoperability. Semantic interoperability refers to the ability for machines to interpret the meaning of words or data from heterogeneous sources, enabling seamless and automated information exchange \cite{Heflin2000}. Conventional approaches to achieving semantic interoperability require datasets to be accompanied by metadata, preferably conforming to an established standard. 

When metadata are available, it can then be indexed and searched to make datasets discoverable within a system. When metadata are created according to different standards or schemes, heterogeneous metadata elements can be mapped to a common standard through a process known as crosswalking \citep{Chan2006, Eric2025}. When different entities use conflicting terms or descriptions, crosswalking can function as a translation layer by mapping local terms to corresponding concepts within a Knowledge Organization System (KOS), such as a controlled vocabulary, thesaurus, or ontology \citep{Chan2006, Zeng2008, Kempf2014}.

However, operationalizing this metadata-centered approach presents several challenges. First, metadata may be incomplete or use non-standardized elements and values. In such cases, metadata must be supplemented or transformed, either through human curation or AI-assisted methods, both of which require an established standard or framework to guide the process. Second, metadata may be entirely absent. Datasets can exist without accompanying documentation, requiring metadata to be created from scratch by data creators or providers to describe the contents and meaning of the data. Finally, relevant metadata standards may be unavailable, incompletely developed, contested, or poorly suited to the needs of the diverse stakeholders who produce and use livestock data. Collectively, these limitations create a disconnect not only in the reuse of data, but also in its initial discovery, as interoperability challenges can prevent users from determining what data exist and how they can be accessed.

Addressing these limitations through conventional metadata practices requires substantial human curation, financial resources, and domain-specific expertise. Moreover, in a fragmented data environment, users must first locate individual resources before determining whether metadata exist, which standards have been applied, and what resources and methods are required to improve their interoperability. Within the livestock agriculture domain, the lack of metadata, agreed-upon metadata standards, and need for geographical and age/sex/production granularity has been demonstrated. Given the urgent need for interoperable and findable data, relying solely on the future development, agreement, and implementation of comprehensive standards is insufficient. The data are needed now.

This paper therefore proposes a \textbf{bottom-up approach} to developing a domain-specific vocabulary directly from livestock data. Rather than relying exclusively on externally defined metadata standards or requiring metadata to be created before data can be discovered, we investigate the use of data elements contained within datasets as a source of descriptive information for dataset discovery and semantics. These terms can be extracted and normalized to produce a vocabulary grounded in the language already present in livestock datasets. We demonstrate that this approach can provide benefits beyond the creation of minimal descriptive metadata, offering a practical pathway toward improving the discoverability and interoperability of heterogeneous livestock data while minimizing the need for prior human curation or AI-assisted metadata generation. 

The objective of this study is to determine whether livestock reporting terminology can be extracted from heterogeneous statistical datasets to construct a bottom-up vocabulary that captures the granularity and variation of real-world livestock classifications. Specifically, the following research questions are addressed: 

\begin{enumerate}
    \item What cattle reporting terminology is present across heterogeneous national and intergovernmental sources?
    \item What patterns of age, sex, production, negation, and other modifiers characterize real-world cattle reporting categories?
    \item How does the resulting vocabulary differ in granularity and coverage from an established agricultural KOS such as AGROVOC? 
\end{enumerate}

By addressing these questions, this study demonstrates how bottom-up vocabulary development can complement top-down KOSs. Rather than requiring terminology to conform to a pre-defined standard before it can contribute to discovery and interoperability, the vocabulary captures information that is already embedded within reporting systems. This provides the empirical basis required for identifying differences in classification granularity, regional terminology, and semantic coverage across livestock sources while providing a foundation for subsequent vocabulary alignment with top-down sources and automated semantic mapping.

\section{Methods}

\begin{figure*}[t]
    \centering
    \includegraphics[width=\textwidth]{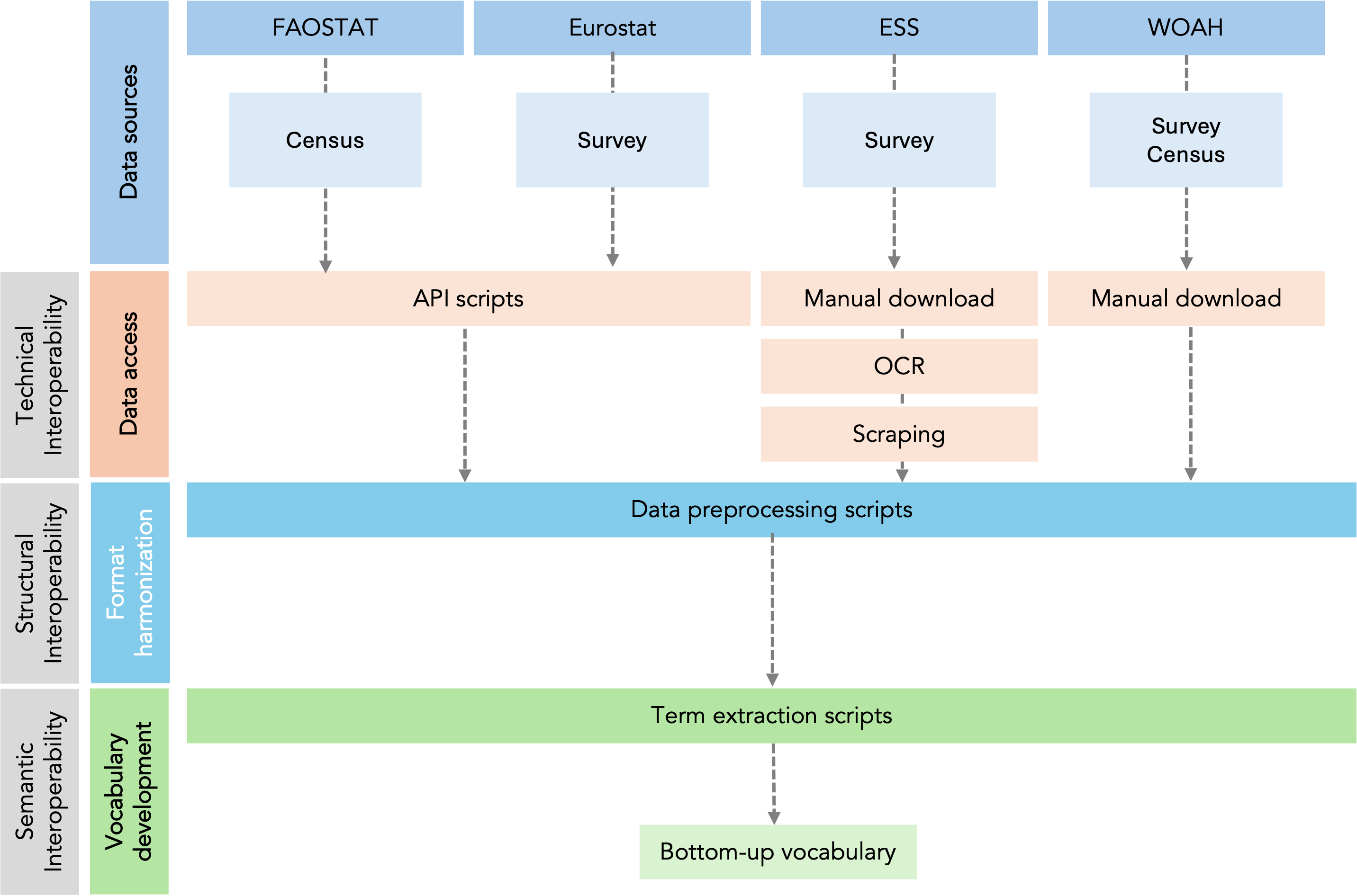}
    \caption{Data extraction process and applications of vocabulary development. FAOSTAT = Food and Agriculture Organization of the United Nations Corporate Statistical Database; ESS = Ethiopia Statistics Service; WOAH = World Organization for Animal Health; API = Application Programming Interface; OCR = Optical Character Recognition. Style of figure adapted from \cite{Kebede2024}}
    \label{fig:data_extraction_pipeline}
\end{figure*}

The methods consist of acquiring cattle population data from livestock data sources by programmatically establishing technical interoperability, harmonizing the format of acquired data sources (achieving structural interoperability), extracting reporting terms from the datasets, extracting terms from a representative agricultural KOS, normalizing all terms, and mapping the extracted terms to the KOS. Figure~\ref{fig:data_extraction_pipeline} provides an overview of the methodology. All scripts used in this study will be made available upon publication.

\subsection{Data Acquisition}

Livestock population data was acquired from four sources: the Food and Agriculture Organization of the United Nations Corporate Statistical Database (FAOSTAT) \cite{FAOSTAT2026}, Ethiopia Statistics Service (ESS) \cite{EthiopiaStat2024}, Eurostat \cite{Eurostat2024}, and the World Organization for Animal Health (WOAH). The data sources are representative of datasets from international organizations (FAOSTAT, WOAH, Eurostat) and a national statistics agency (ESS). Two datasets were gathered from the FAO (the Emissions: Enteric Fermentation (dataset code: GE) and Crops and Livestock Products (dataset code: QCL) datasets). Table~\ref{tab:data-sources} summarizes the data sources, access methods, and collection dates. To acquire the data, technical interoperability was achieved. 

\subsection{Technical Interoperability}

Technical interoperability was achieved through tailoring the data acquisition method to the machine-readability of the data. Where Application Programming Interfaces (APIs) were available, Python scripts were developed to access data programmatically, enabling a reproducible data pipeline. For sources lacking APIs, data was downloaded manually from official data portals. In the case of the ESS, data was available in PDF reports. Optical Character Recognition (OCR) and scraping were used to transform PDF data into machine-readable datasets.

\begin{table*}[t]
\centering
\footnotesize
\setlength{\tabcolsep}{3pt}
\renewcommand{\arraystretch}{1.15}
\begin{tabularx}{\textwidth}{@{}>{\raggedright\arraybackslash}p{3.6cm} >{\raggedright\arraybackslash}p{3.8cm} >{\raggedright\arraybackslash}X >{\raggedright\arraybackslash}p{2.6cm}@{}}
\toprule
\textbf{Data source} & \textbf{Dataset / table} & \textbf{Link to data or API call} & \textbf{Acquisition method} \\
\midrule
\multirow[t]{2}{3.6cm}{Food and Agriculture Organization of the United Nations Statistical Database} & Emissions: Enteric Fermentation (GE) & \url{https://fenixservices.fao.org/faostat/static/bulkdownloads/Emissions_Agriculture_Enteric_Fermentation_E_All_Data.zip} & Direct download \\
\cmidrule(lr){2-4}
 & Crops and livestock products (QCL) & \url{https://fenixservices.fao.org/faostat/static/bulkdownloads/Production_Crops_Livestock_E_All_Data.zip} & Direct download \\
\midrule
World Organisation for Animal Health & Livestock population & \url{http://gbadske.org:9000/GBADsLivestockPopulation/oie?species=*&year=\%3Cyear\%3E&format=file} & GBADs API \\
\midrule
\multirow[t]{5}{3.6cm}{Eurostat} & apro\_mt\_lscatl& \url{https://ec.europa.eu/eurostat/api/dissemination/statistics/1.0/data/apro_mt_lscatl?format=JSON} & Eurostat API \\
\midrule
Ethiopia Central Statistics Agency (EthCSA) & Ethiopia Annual Agriculture Sample Survey: Report on Livestock and Livestock Characteristics & \url{https://github.com/GBADsInformatics/pdfScrapingETHCSA/data} & GBADs GitHub \\
\bottomrule
\end{tabularx}
\caption{Data sources and acquisition methods. All data was collected in February 2026.}
\label{tab:data-sources}
\end{table*}
\clearpage

\subsection{Structural Interoperability}

Structural interoperability was achieved through the development of dataset-specific scripts used to harmonize heterogeneously formatted data. Format harmonization involved uniformly structuring columns and rows, and standardizing column headers to support further processing. Once the format of the data was harmonized, a pipeline was created that could automatically extract terms from datasets. 

\subsection{Term Extraction}

Terms were extracted from the datasets using the pipeline mentioned above, resulting in a vocabulary of livestock terms used in real-world reporting practices. For the scope of this study, all terms related to cattle were retained for processing and normalization, while other livestock terms were excluded. Due to differences in production and utility, other species of livestock are expected to require an alternative set of normalization rules. This is discussed in more detail later in the paper.

\subsection{Reference Vocabulary Extraction}

Reference terms were extracted from AGROVOC. AGROVOC was selected due to its relevance to agricultural terminology. Published as Linked Open Data (LOD), it is a FAIR vocabulary that is managed by the FAO and is the largest agricultural multilingual vocabulary in the world \cite{Subirats2022}. Therefore, mapping terms to AGROVOC provides a unified vocabulary layer to an otherwise heterogeneously described landscape of livestock terms. 

\begin{figure}
    \centering
    \includegraphics[width=0.4\linewidth]{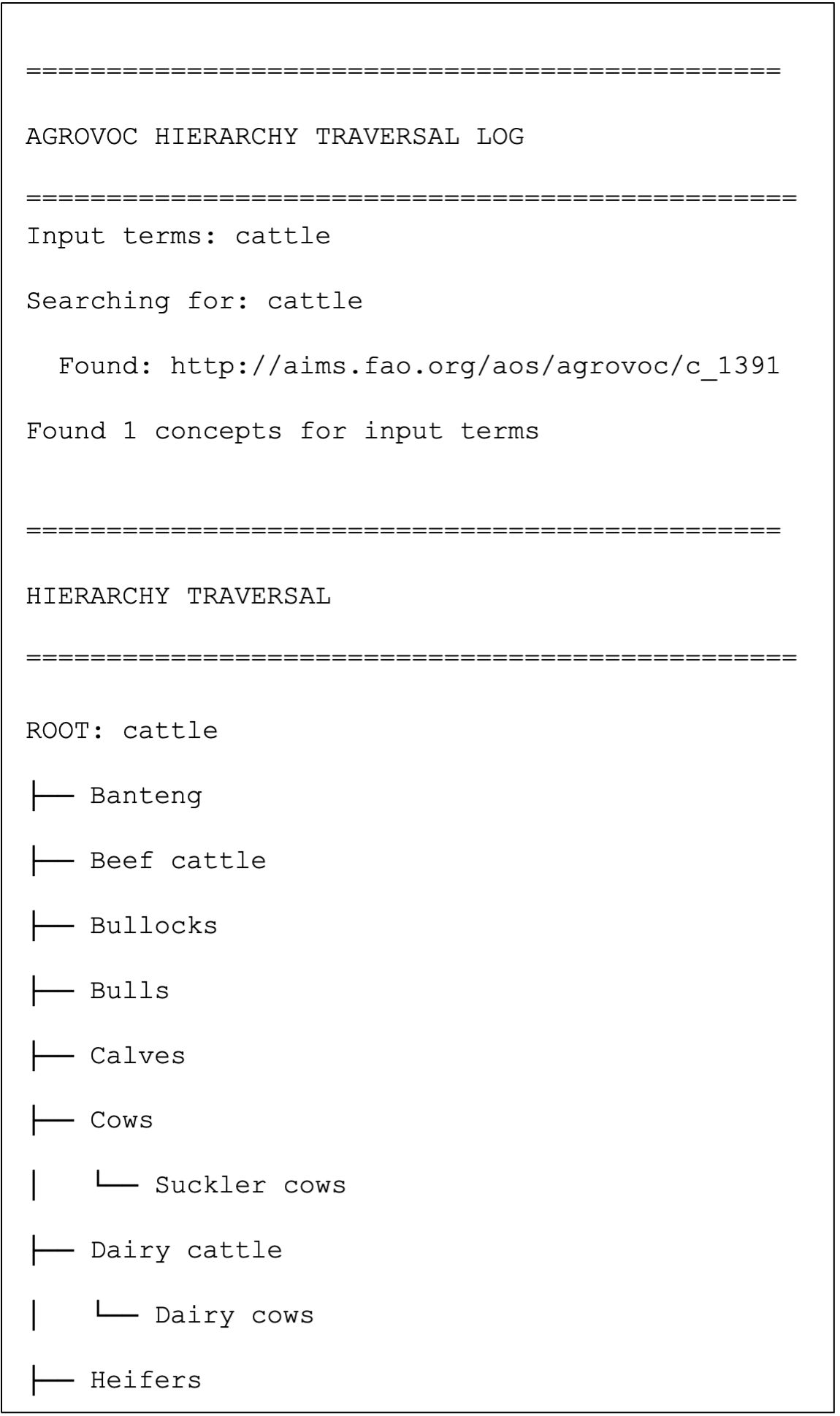}
    \caption{Recursive tree traversal log for term ``cattle'' (c\_1391) from AGROVOC.}
    \label{fig:recursive_tree}
\end{figure}

English AGROVOC terms were extracted using a Python script developed to interact with the AGROVOC API. Using \texttt{cattle (c\_1391)} as the root concept for input, a recursive tree traversal was used to discover child terms. As shown in Figure~\ref{fig:recursive_tree}, the script discovered child terms of the root concept \texttt{cattle} by traversing the AGROVOC tree using the \texttt{skos:narrower} relationship. The root concept \texttt{cattle} was selected by navigating the SKOS hierarchy tree on the AGROVOC website, ensuring that the most relevant top-level concept was used. Each run of the Python script generated a traversal log, ensuring provenance, transparency, and reproducibility in the vocabulary extraction process. The term list was manually reviewed to ensure that all terms gathered from the API were relevant to cattle. 

\subsection{Normalization}

\begin{figure}[b]
    \centering
    \includegraphics[width=0.5\linewidth]{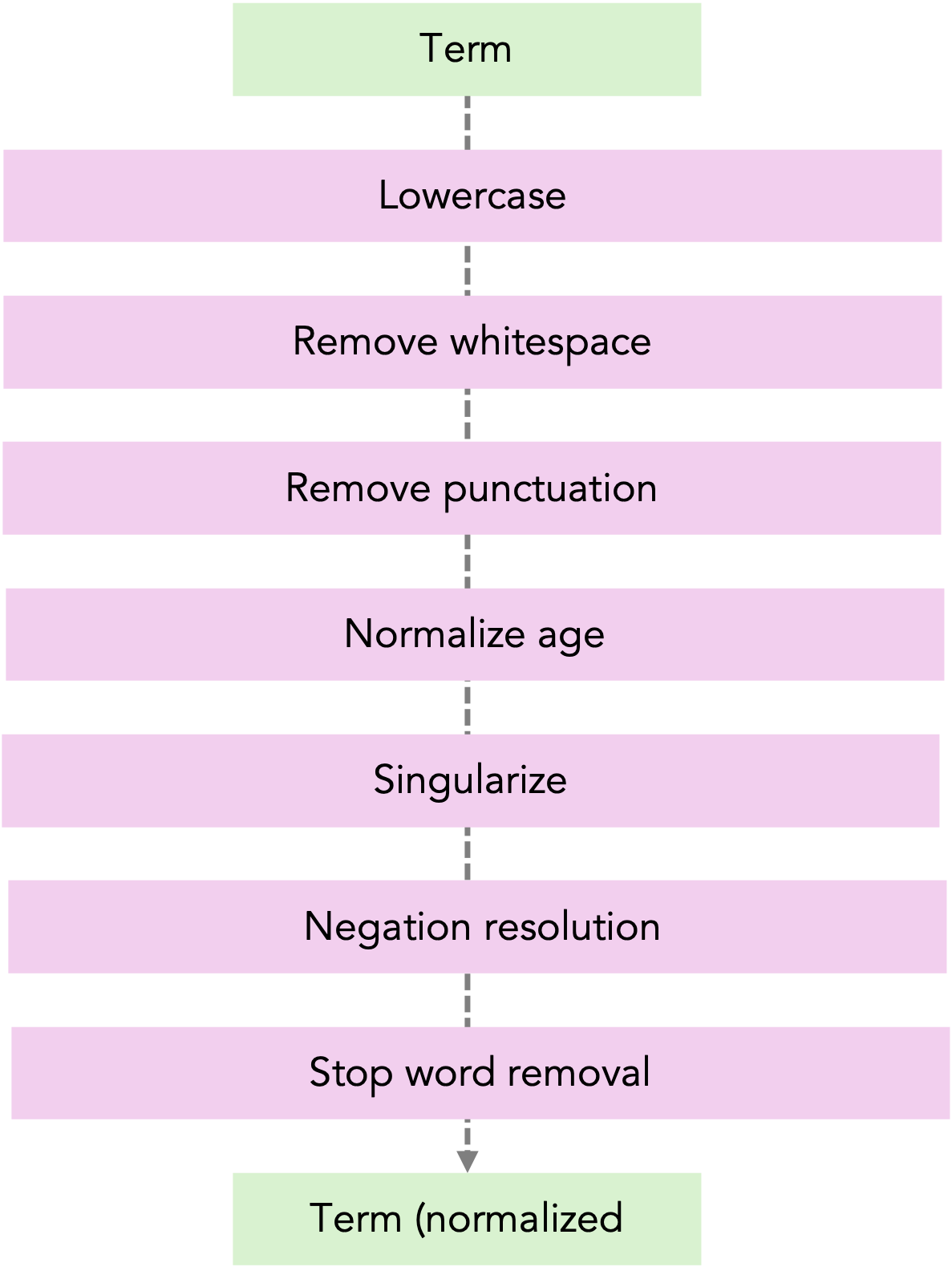}
    \caption{Text normalization pipeline.}
    \label{fig:text_normalization_pipeline}
\end{figure}

Terms from the datasets and AGROVOC were normalized in a text-preprocessing workflow, which was implemented as a pipeline in Python (Figure~\ref{fig:text_normalization_pipeline}). Text normalization is a critical step in many Natural Language Processing (NLP) tasks, and allow for analysis to identify patterns and similarities in the semantic composition of a term \cite{Aliero2023}.

First, all terms were converted to lowercase and trailing and leading whitespace and punctuation were removed. Raw terms were analyzed to determine the structure of the term. Rule-sets were created to normalize variations in how terms were expressed. Examples are provided in the results section. Next, TextBlob \cite{Loria2014} was used to normalize plural words to a singular form (e.g., \textit{calves} to \textit{calf}). NLTK \cite{Bird2009} and spaCy \cite{Honnibal2020} were also trialed for stemming and lemmatization, but they did not consistently handle irregular plural forms, often failing to resolve to the correct root. For example, rather than inflecting \textit{calves} to \textit{calf}, the resolved form was \textit{calv}.

Age-related categories were then standardized to consolidate variations in how age categories were expressed and to preserve semantic meaning. In livestock reporting, terms such as \textit{and}, \textit{less}, and \textit{than} are used to assert age ranges (e.g. \textit{female cattle 1 year \texttt{and} \texttt{less} \texttt{than} 3 years}). Therefore, standardization of terms needs to occur before the removal of standard stopwords. For example, \textit{female cattle 1 year and less than 3 years} would become \textit{female cattle 1 year 3 years}. Terms were then analyzed to create a set of rules to normalize age ranges. Rules are provided in the results section since they are a product of the analysis. Ages reported in months were converted to years. For example, \textit{female cattle 6 months and less than 1 year} was normalized to \textit{female cattle age\_0.5\_to\_1\_year}. Once age patterns were identified and rules were created, spaCy’s EntityMatcher \cite{Honnibal2020} was used to identify these patterns and replace them with the standardized labels.

To normalize negated terms, negation cues specific to livestock terms were identified such as \textit{not} and \textit{not for}. These were resolved to a \texttt{NOT\_} prefix, normalizing \textit{non-dairy} to \texttt{NOT\_dairy} and \textit{not for slaughter} to \texttt{NOT\_slaughter}. 

Finally, stopwords were removed using NLTK’s English corpus. Additional custom stopwords included \textit{total} and \textit{live}. For example, \textit{total beef cattle} was normalized to \textit{ beef cattle} and \textit{live bovine animal} to \textit{bovine animal}, respectively.

Output files from the pipeline provide columns for each stage in the normalization process, ensuring research is reproducible and data provenance can be traced.

\section{Results and Discussion}

\subsection{Terminology and Modifiers that Characterize Real-World Cattle Root Terms}

Across all datasets, 63 unique cattle terms exist. Like nomenclature used in medical coding and terminology, terms were composed of a root term with modifiers \cite{Hirsch2026}. Modifiers include prefixes, suffixes or other modifications to a root word that alter, clarify, or specify details about the context without changing the meaning of a word. For example, when red is a colour modifier for the root term truck (i.e. red truck), red specifies the colour of the truck without changing the fact that we are describing a truck. 

\begin{figure}[p]
    \centering
    \makebox[\textwidth][c]{%
        \includegraphics[width=1.5\textwidth]{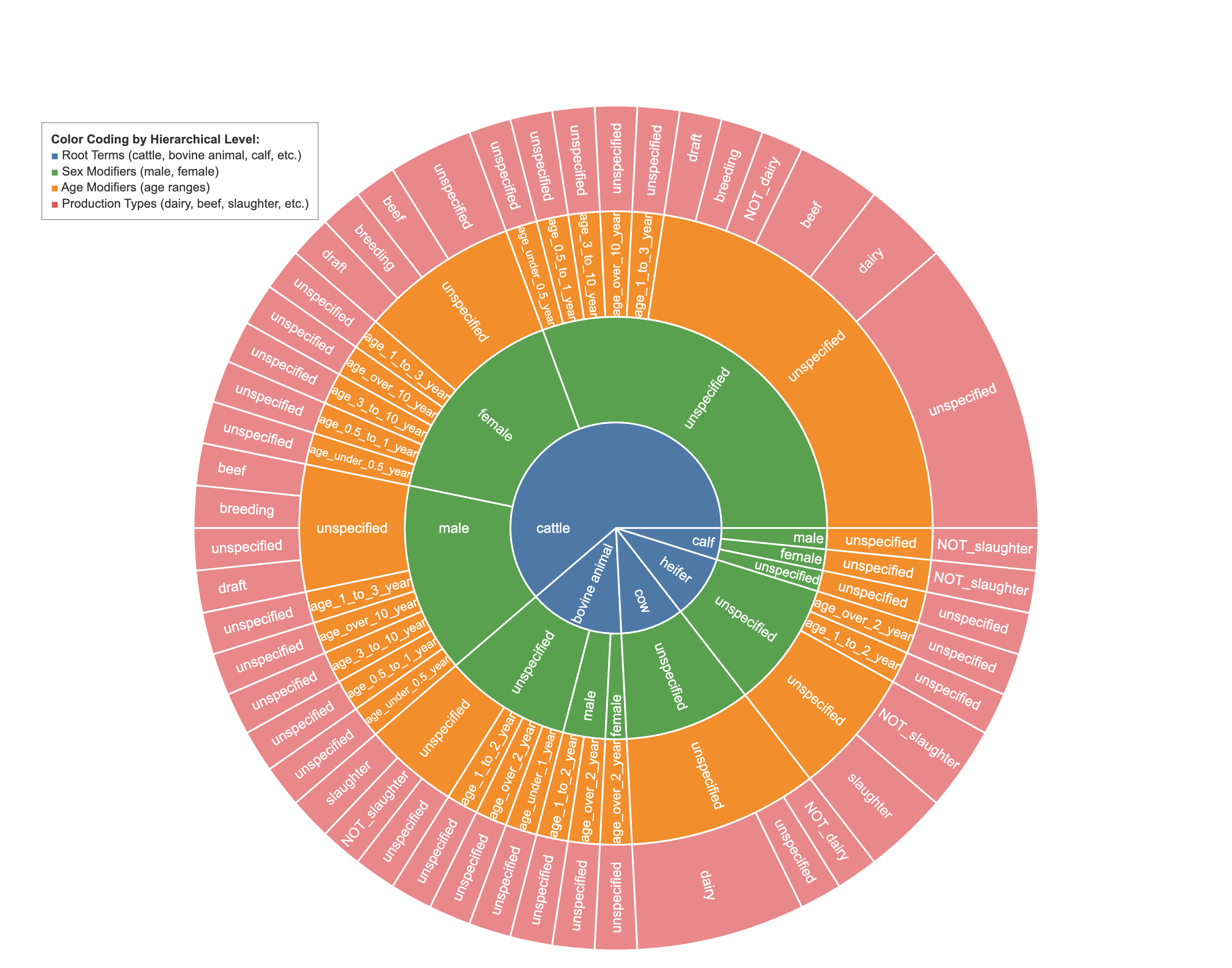}%
    }
    \caption{Sunburst chart showing the hierarchical structure of livestock reporting categories occurring in real-world reporting categories.}
    \label{fig:sunburst}
\end{figure}

Cattle terms included age, sex, and production modifiers, which provide additional context to the root term. By identifying the modifiers, rule-sets were created to normalize the terms and modifiers based on their type.  Table~\ref{tab:composition} provides the root terms and sex, age, and production modifiers that terms are composed of. The terms themselves consisted of a combination of a root term with one or more of the modifiers. Figure~\ref{fig:sunburst} shows the hierarchical structure of the composition of a cattle term. 

The composition of the cattle term, consisting of a root term with modifiers, reflects institutional priorities and real-world information needs. Reporting categories are not chosen arbitrarily. For example, in Canada, ``content consultations with data users across federal government departments and provincial ministries, agricultural associations, and educational institutions'' are held to inform questions and classifications used in the Census of Agriculture \cite{StatCan2025}. These consultations demonstrate that questionnaires (and thus reporting categories) are chosen based on the real-world information needs of the individuals or organizations who use the data.

\begin{table}[b]
    \begin{tabularx}{\textwidth}{@{}>{\raggedright\arraybackslash}p{3.2cm}X@{}}
        \toprule
        \textbf{Component} & \textbf{Values} \\
        \midrule
        Root terms & \textit{cattle}, \textit{bovine animal}, \textit{cow}, \textit{heifer}, \textit{calf} \\
        Sex modifiers & female, male \\
        Age modifiers & age\_1\_to\_3\_year, age\_over\_10\_year, age\_3\_to\_10\_year, age\_0.5\_to\_1\_year, age\_under\_0.5\_year, age\_1\_to\_2\_year, age\_over\_2\_year, age\_under\_1\_year, and age\_1\_year \\
        Production modifiers & draft, breeding, beef, dairy, NOT\_dairy, slaughter, NOT\_slaughter \\
        \bottomrule
    \end{tabularx}
    \caption{Composition of livestock terms collected from the Ethiopia Statistical Service (ESS), FAOSTAT, Eurostat, and the World Organization for Animal Health (WOAH).}
    \label{tab:composition}
\end{table}
\FloatBarrier

\clearpage
\clearpage

This logic justifies the use of the reporting categories and their composition as a relevant choice for metadata fields or search functionality. The inclusion of reporting categories in bureaucratic collection therefore implies that the categories reflect the needs of data users. Therefore, the composition of terms demonstrates the inherit utility of a modifier as descriptive metadata and for data querying, search, and retrieval purposes. 

\subsubsection{Sex and Production Modifiers}

Sex information in the cattle term either occurs through a modifier or is inherent in the meaning of the term. In the dataset, sex modifiers directly label terms as either male or female (i.e. \texttt{female} \textit{cattle} \texttt{age\_1\_to\_3\_year}). Implicit sex information in terms occurs when the term suggests the animal is inherently male or female but does not explicitly label the term as male or female. For example, \textit{heifer} implicitly denotes a female animal while a \textit{bull} is male \cite{Evans2007}. 

Production modifiers can also contain implicit sex information, though with varying reliability. For example, the production modifier \texttt{dairy} implies a female animal since milk production is a biologically female trait. However, male calves born to dairy cattle may be kept for breeding purposes and are referred to as male dairy cattle. For example, in Canada, the population of male dairy calves is reported by Agriculture and Agri-Food Canada \cite{StatCan2025}. 

Production modifiers such as \texttt{beef} are ambiguous in sex. A \texttt{beef} cattle in a high-intensive production system is typically male. However, heifers (female) may be first used for dairy production and later be slaughtered for meat (beef) production. For example, results show that a \textit{heifer} occurs with both \texttt{slaughter} and \texttt{NOT\_slaughter}, indicating heifers may be used for meat production after they are slaughtered. Therefore, the modifier, \texttt{beef}, alone cannot reliably imply biological sex without additional context.

Deciphering sex information from terms is useful for automated metadata generation and data discovery. Tagging terms, and therefore datasets, according to sex enables search by sex in data retrieval systems. For example, by encoding \texttt{female} with the term \textit{cow}, systems can allow users to search specifically for \texttt{female cattle data}. Terms should be flagged for manual intervention if systems cannot reliably extract sex modifiers to determine whether questionnaires or data collection forms explicitly express sex.


\subsubsection{Age Modifiers}

Root terms are sometimes accompanied by age ranges that provide information about the age of the animal being reported. Four distinct age rules were established were present across all terms. To normalize terms, the following rule-sets were created (Table~\ref{tab:age_modifiers}): 

\begin{enumerate}
    \item age range
    \item discrete age
    \item minimum age range
    \item maximum age range 
\end{enumerate}

\begin{table}[b]
    \centering
    \footnotesize
    \setlength{\tabcolsep}{3pt}
    \renewcommand{\arraystretch}{1.15}
    \begin{tabularx}{\textwidth}{@{}>{\raggedright\arraybackslash}p{2.2cm} >{\raggedright\arraybackslash}X >{\raggedright\arraybackslash}p{3.0cm} >{\raggedright\arraybackslash}p{3.0cm}@{}}
        \toprule
        \textbf{Rule} & \textbf{Example} & \textbf{Age Modifier} & \textbf{Standard Age Modifier} \\
        \midrule
        Age range & Female Cattle 1 year and less than 3 years & X year(s) and less than Y year(s) & age\_1\_to\_3\_year\\
        Discrete age & Heifers, 1 year old, for slaughter & X year(s) old & age\_1\_year \\
        Minimum age range & Female Cattle less than 6 months & less than X month(s) & age\_less\_0.5\_year \\
        Maximum age range & Female Cattle 10 years and older & X year(s) and older & age\_over\_10\_year \\
        \bottomrule
    \end{tabularx}
    \caption{Age modifier rules, examples, and standards used to normalize age modifiers in livestock terms.}
    \label{tab:age_modifiers}
\end{table}

These rules were developed to consolidate heterogeneous forms while preserving the information contained in reported age categories. The standardized representation replaces spaces with underscores to provide a consistent machine-readable label and to distinguish the components of the age modifier.

The \textbf{age range rule} is applied when an age category contains both a lower and upper bound. For example, \textit{female cattle 1 year and less than 3 years} is normalized to \texttt{age\_1\_to\_3\_year}. Similarly, \textit{heifers, 1 to less than 2 years old} is normalized to \texttt{age\_1\_to\_2\_year}. The \textbf{discrete age rule} is applied when ages are presented at a set time. The only example of this in the dataset used is \textit{Heifers, 1 year old, for slaughter}, being normalized to \texttt{age\_1\_year}. The \textbf{minimum age range} rule is applied when an age category specifies an upper bound. For example, \textit{less than 6 months} is normalized to \texttt{age\_less\_0.5\_year}.  Conversely, the \textbf{maximum age range} rule is applied when a category specifies a lower bound without an upper bound. For example, \textit{10 years and older} is normalized to \texttt{age\_over\_10\_year}.

These transformations allow heterogeneous representations of age to be represented using a common vocabulary. This distinction is important because the objective of normalization is not to remove age information, but to separate the age component from variation in how reporting agencies express that information. The resulting age modifiers enable the creation of reusable components of reporting categories rather than as independent, dataset-specific strings. The generation of age-related metadata from the terms will allow for data retrieval systems to support querying by age. For example, a drop down menu or choice of age categories can be offered, allowing users to filter datasets by age. This is particularly useful as age-related biological processes in herds possess different methane emissions, nutritional outputs, and can be sold at different prices. 

As additional datasets are incorporated into the vocabulary, the rule set may need to be expanded to accommodate different ways age may be expressed. Because Ethiopia was the only national source included in this study, additional age categories may be identified when data from other national reporting systems are incorporated. The resulting vocabulary should therefore be considered extensible rather than exhaustive.

The normalized age modifiers also provide a potential bridge between reporting categories and qualitative age categories used in livestock and climate modelling. For example, the GBADs Animal Health Loss Envelope (AHLE) uses qualitative categories such as ``adult'', ``sub-adult'', and ``juvenile'' \cite{Gilbert2024}. Creating mappings between these categories and the observed age ranges could enable model inputs to be derived from reported data. Maintaining the provenance would allow analysts and modellers to identify the original age categories underlying model inputs, compare alternative mappings, and assess the sensitivity of model outputs to different age classifications while allowing for these analyses to be more reproducible. 

\subsection{Comparison to AGROVOC}

Text based comparison to AGROVOC included determining whether terms in AGROVOC also occurred in reporting terms. The normalized form of the AGROVOC terms retrieved included: banteng, beef cattle, bullock, bull, calf, cow, dairy cattle, dairy cow, heifer, suckler cow, yearling bull, and zebu. Of the AGROVOC terms retrieved, the following terms never matched to any of the terms in the bottom-up vocabulary: banteng, bullock, bull, suckler cow, yearling bull, and zebu. Table~\ref{tab:num_agrovoc_matches} shows the number of matches from the real-world cattle reporting terms to the normalized AGROVOC terms. 

\begin{table}[t]
    \begin{tabularx}{\textwidth}{@{}>{\raggedright\arraybackslash}p{3.4cm}X>{\raggedleft\arraybackslash}p{1.3cm}@{}}
        \toprule
        \textbf{AGROVOC term (normalized)} & \textbf{Matched reporting terms} & \textbf{Matches} \\
        \midrule
        cattle & cattle; cattle NOT\_dairy; male female cattle; female beef cattle; female cattle age\_1\_to\_3\_year; female cattle age\_over\_10\_year; female cattle age\_3\_to\_10\_year; female cattle age\_0.5\_to\_1\_year; female cattle breeding; female cattle purpose; female cattle age\_under\_0.5\_year; female draft cattle; male beef cattle; male cattle age\_1\_to\_3\_year; male cattle age\_over\_10\_year; male cattle age\_3\_to\_10\_year; male cattle age\_0.5\_to\_1\_year; male breeding cattle; male cattle purpose; male cattle age\_under\_0.5\_year; male draft cattle; beef cattle; cattle age\_1\_to\_3\_year; cattle age\_over\_10\_year; cattle age\_3\_to\_10\_year; cattle age\_0.5\_to\_1\_year; cattle breeding; cattle purpose; cattle age\_under\_0.5\_year; draft cattle; fat cattle; hide cattle fresh; meat cattle; offal edible cattle; cattle buffalo & 35 \\
        No AGROVOC match & bovine animal age\_1\_to\_2\_year; bovine animal age\_over\_2\_year; bovine animal age\_under\_1\_year; bovine animal age\_under\_1\_year slaughter; bovine animal age\_under\_1\_year NOT\_slaughter; female bovine animal age\_over\_2\_year; bovine animal; male bovine animal age\_1\_to\_2\_year; male bovine animal age\_over\_2\_year & 9 \\
        heifer & heifer age\_1\_to\_2\_year; heifer age\_1\_year slaughter; heifer age\_1\_year NOT\_slaughter; heifer age\_over\_2\_year; heifer age\_over\_2\_year slaughter; heifer age\_over\_2\_year NOT\_slaughter & 6 \\
        cow & cow gave milk last 12 month; dairy cow; milk whole fresh cow; cow; NOT\_dairy cow & 5 \\
        beef cattle & adult beef cattle; female beef cattle; male beef cattle; beef cattle & 4 \\
        calf & calf; female calf age\_under\_1\_year NOT\_slaughter; male calf age\_under\_1\_year NOT\_slaughter & 3 \\
        dairy cattle & cattle dairy; adult dairy cattle & 2 \\
        dairy cow & dairy cow & 1 \\
        \bottomrule
    \end{tabularx}
    \caption{Number of matches from real-world cattle reporting terms and normalized AGROVOC cattle terms.}
    \label{tab:num_agrovoc_matches}
\end{table}

Across the terms extracted from the datasets, AGROVOC terms provide a generalized vocabulary that map directly to root terms and in some cases, production modifiers, when explicitly expressed in the AGROVOC term. For example, beef cattle, cows, heifers, dairy cattle, and dairy cows were represented in AGROVOC directly and were therefore matched. Other combinations of age, sex, and production modifiers which were not expressed in AGROVOC, matched more generally to \textit{ cattle}. For example, reporting terms containing age categories, sex categories, or production characteristics frequently matched \textit{cattle} rather than a correspondingly specific AGROVOC concept. Conversely, the root term \textit{bovine} did not match any of the captured AGROVOC terms.

This demonstrates a lack of granularity relative to the cattle terms represented in the reporting categories used by the datasets. Although AGROVOC represents broad concepts that map to many of the terms, it does not represent all of the age, sex, or production categories encoded in those terms. Consequently, mapping reporting terms to AGROVOC may result in information loss, particularly where specific terms are mapped to the broader concept \textit{cattle}. This may limit the utility of AGROVOC for practice-relevant interoperability and information retrieval where distinctions beyond root terms and a limited number of production modifiers are required. For example, if AGROVOC terms are used to support data search and findability, users will be unable to retrieve datasets based on specific age categories when these distinctions are not represented in the corresponding AGROVOC concepts. This is potentially important for livestock population data used as inputs to climate and economic models, where age-disaggregated livestock populations may be required.

This finding is consistent with the description of AGROVOC; AGROVOC is defined as a general-purpose vocabulary designed to cover concepts under FAO’s areas of interest to provide data access and visibility across domains \cite{FAO_AGROVOC_2026}. Therefore, while AGROVOC remains an important tool for interoperability, the results suggest that bottom-up terms contain information that is not currently represented in AGROVOC. As AGROVOC is currently the world's largest agricultural vocaulary, future work should consider additions to AGROVOC to promote its utility. Alternative vocabularies, such as USDA National Agricultural Thesaurus (NALT) \cite{USDA_NALT_2022}, can also be explored to determine whether specialized terms are more effectively captured. 

Embedding-based models may provide additional support for crosswalking between terms that are semantically similar but not lexically equivalent; for example, \texttt{meat} occurs in reported terms while \texttt{beef} is present in the AGROVOC term \textit{beef cattle}. Mappings between terms containing \texttt{meat} and \texttt{beef} would allow meat cattle datasets to be retrieved when using the term \textit{beef cattle} to search for data. Future work should consider whether general-purpose models are effective in creating such mappings or if other approaches are needed.

Beyond considerations of the use of standards such as AGROVOC in the data, these findings are relevant to the application of the FAIR guiding principles. Although insufficient metadata can limit the reusability of datasets by providing inadequate context regarding how data were collected, the purposes for which they were collected, and what the data represent, the application of standards alone does not necessarily resolve these limitations. The results show that standards and controlled vocabularies have their own scope and level of granularity which can constrain the information represented in metadata. As demonstrated by the AGROVOC comparison, mapping more specific reporting terms to broader concepts may support interoperability at a general level while simultaneously resulting in information loss that affects more specific forms of findability and reuse. Future work should therefore consider not only whether standards are applied, but also the limitations of the standards themselves and their implications for the findability, interoperability, and reusability of data.

Finally, as categories change between countries and by data source, it is important to ensure that cultural-specific categories are exposed in metadata. Ultimately, the inclusion of alternative categories would ensure that all groups could search for terms according to their own specific needs. 

\subsection{Conclusion}

This study shows that the composition of reporting categories used in livestock population data provide semantic structure. The rule-based, bottom-up extraction and normalization of terms results in a vocabulary that allows the semantics of the terms to be leveraged. The composition of term captures granularity that is not present in AGROVOC but that is desired by researchers \cite{Kebede2024}. The vocabulary can be used to improve the interoperability and findability of disparate and heterogeneous livestock data. However, it does not replace the need for contextual and descriptive metadata required for the reusability of datasets themselves. 

The research presented allows datasets to be made findable and interoperable before waiting for metadata to be generated and standards to be created, providing a pragmatic approach that emphasizes the nuance of what the requirements of findability and interoperability are in this context. The approach provides a faster and less-resource intensive solution, providing a pathway to improving the discoverability and interoperability of cattle data without additional human or AI-assisted curation or metadata generation. We argue that the approach provided to recover the semantics from data and provide terms that can be used to facilitate the findability of data is both novel and necessary to move forward and challenge metadata-only or FAIR-only thinking. Indeed this argument aligns with \cite{Villa2017}'s view that ``approaches that can facilitate the development of consistent semantics beyond textual metadata and controlled vocabularies become essential.'' 

Furthermore, the identification and normalization of modifiers provides a unique opportunity to improve the standardization of legacy datasets; top-down standardization of metadata can result in information loss, as demonstrated by the comparison from the bottom-up vocabulary to AGROVOC. However, since reporting categories are expected to change over time, it is important to be able to recover data from past years. Bottom-up vocabularies provide such an approach.

Future work should collect and extract data from additional national sources to make the normalization rule-set and vocabulary more robust. The approach used can also be applied to other species of livestock to generate rules and identify modifiers that are species-specific. 

Additionally, the co-occurrence of modifiers as combinations of some modifiers could be logically incorrect. For example, in the dataset collected in this study, the root term, calf, always occurs with the age modifier \texttt{age\_under\_1\_year}. \textit{Calf} implies a young animal and therefore should never be modified with an age range that represents a juvenile or adult animal. An evaluation of such co-occurrence networks of modifiers can generate a set of rules used for reasoning, error-handling, and data quality that would not be possible with existing top-down standards.

\section{Acknowledgments}

We gratefully acknowledge the support of the Natural Sciences and Engineering Research Council of Canada (NSERC), funding reference number [570288-2022].

\bibliographystyle{plainnat}
\bibliography{refs}

@article{Aliero2023,
  title={Systematic review on text normalization techniques and its approach to non-standard words},
  author={Aliero, Abubakar Ahmad and Bashir, Sulaimon Adebayo and Aliyu, Hamzat Olanrewaju and Tafida, Amina Gogo and Bashar, Umar Kangiwa and Nasiru, Muhammad Dankolo},
  year={2023},
  journal={International Journal of Computer Applications}
}

@book{Bird2009,
  author    = {Bird, Steven and Klein, Ewan and Loper, Edward},
  title     = {Natural Language Processing with {Python}: Analyzing Text with the {Natural Language Toolkit}},
  year      = {2009},
  publisher = {O'Reilly Media, Inc.},
  isbn      = {978-0596516499}
}

@article{Cabrera2025,
  title={Data integration and analytics in the dairy industry: challenges and pathways forward},
  author={Cabrera, Victor E and Bewley, Jeffrey and Breunig, Mitch and Breunig, Tom and Cooley, Walt and De Vries, Albert and Fourdraine, Robert and Giordano, Julio O and Gong, Yijing and Greenfield, Randall and others},
  journal={Animals},
  volume={15},
  number={3},
  pages={329},
  year={2025},
  publisher={MDPI}
}

@article{Chan2006,
  title={Metadata interoperability and standardization--a study of methodology part {I}},
  author={Chan, Lois Mai and Zeng, Marcia Lei},
  journal={D-Lib magazine},
  volume={12},
  number={6},
  pages={1082--9873},
  year={2006},
  publisher={Corporation for National Research Initiatives}
}

@inproceedings{Eric2025,
  title={Implementing interoperability-Metadata Schema and Crosswalk Registry approach to {FAIR} metadata mappings},
  author={Eric, Clarin},
  booktitle={Proceedings of EUNIS},
  volume={105},
  pages={58--68},
  year={2025}
}

@misc{EthiopiaStat2024,
  title = {Our Survey Reports},
  howpublished = {\url{https://www.statsethiopia.gov.et/our-survey-reports/}},
  note = {Accessed: 2024-08-27},
  year = {2024},
  author = {{Ethiopia Central Statistical Agency}}
}

@misc{Eurostat2024,
  title = {Eurostat Data Web Services},
  howpublished = {\url{https://ec.europa.eu/eurostat/data/web-services}},
  note = {Accessed: 2024-08-27},
  year = {2024},
  author = {{Eurostat}}
}

@article{Evans2007,
  title={Beef improvement terminology},
  author={Evans, John and Apple, Ken},
  year={2007},
  publisher={Oklahoma Cooperative Extension Service}
}

@techreport{FAO2025,
  author      = {{Food and Agriculture Organization of the United Nations}},
  title       = {Digital Agriculture and {AI} Innovation Roadmap for the Global Agrifood Systems Transformation},
  institution = {FAO},
  year        = {2025},
  address     = {Rome, Italy},
  url         = {https://openknowledge.fao.org/items/f6e16e80-31d1-45cd-81de-64638d68c836}
}

@book{FAOSTAT2026,
  title = {{FAOSTAT} {Statistics} {Database}},
  url = {https://www.fao.org/faostat/en/#home},
  publisher = {Food and Agriculture Organization},
  author = {{FAO}},
  year = {2026},
  language = {English}
}

@misc{Hirsch2026,
  author       = {Hirsch, Barron},
  title        = {Medical Terms: Prefixes, Roots and Suffixes (Comprehensive List)},
  howpublished = {\url{https://globalrph.com/medical-terms-introduction/}},
  publisher    = {GlobalRPH},
  note         = {Accessed September 4, 2026}
}

@article{Villa2017,
  title={Semantics for interoperability of distributed data and models: Foundations for better-connected information},
  author={Villa, Ferdinando and Balbi, Stefano and Athanasiadis, Ioannis N and Caracciolo, Caterina},
  journal={F1000Research},
  volume={6},
  year={2017}
}

@article{Gilbert2024,
  title={Interpretation and utility of the {A}nimal {H}ealth {L}oss {E}nvelope as part of the {G}lobal {B}urden of {A}nimal {D}iseases analytical process},
  author={Gilbert, W},
  journal={OIE Revue Scientifique Et Technique},
  volume={43},
  pages={108--114},
  year={2024},
  publisher={OIE (World Organisation for Animal Health)}
}

@inproceedings{Heflin2000,
  title={Semantic interoperability on the web},
  author={Heflin, Jeff and Hendler, James},
  booktitle={Proceedings of extreme markup languages},
  volume={2000},
  pages={111--120},
  year={2000}
}

@article{Herrero2013,
  title     =   {Biomass use, production, feed efficiencies, and greenhouse gas emissions from global livestock systems},
  author    =   {Herrero, Mario and Havl{\'\i}k, Petr and Valin, Hugo and Notenbaert, An and Rufino, Mariana C and Thornton, Philip K and Bl{\"u}mmel, Michael and Weiss, Franz and Grace, Delia and Obersteiner, Michael},
  journal   =   {Proceedings of the National Academy of Sciences},
  volume    =   {110},
  number    =   {52},
  pages     =   {20888--20893},
  year      =   {2013},
  publisher =   {National Acad Sciences}
}

@article{Honnibal2020,
  author    = {Honnibal, Matthew and Montani, Ines and Van Landeghem, Sofie and Boyd, Adriane},
  title     = {{spaCy}: Industrial-strength Natural Language Processing in Python},
  year      = {2020},
  doi       = {10.5281/zenodo.1212303},
  publisher = {Zenodo}
}

@article{Jemberu2022,
  title={Population, biomass, and economic value of small ruminants in Ethiopia},
  author={Jemberu, Wudu T and Li, Yin and Asfaw, Wondwosen and Mayberry, Dianne and Schrobback, Peggy and Rushton, Jonathan and Knight-Jones, Theodore JD},
  journal={Frontiers in veterinary science},
  volume={9},
  pages={972887},
  year={2022},
  publisher={Frontiers Media SA}
}

@article{Huntington2021,
   author = {B. Huntington and T. M. Bernardo and M. Bondad-Reantaso and M. Bruce and B. Devleesschauwer and W. Gilbert and D. Grace and A. Havelaar and M. Herrero and T. L. Marsh and S. Mesenhowski and D. Pendell and D. Pigott and A. P. Shaw and D. Stacey and M. Stone and P. Torgerson and K. Watkins and B. Wieland and J. Rushton},
   doi = {10.20506/RST.40.2.3246},
   issn = {0253-1933},
   issue = {2},
   journal = {Revue scientifique et technique (International Office of Epizootics)},
   month = {8},
   pages = {567-584},
   pmid = {34542092},
   publisher = {Rev Sci Tech},
   title = {{Global Burden of Animal Diseases}: a novel approach to understanding and managing disease in livestock and aquaculture},
   volume = {40},
   url = {https://pubmed.ncbi.nlm.nih.gov/34542092/},
   year = {2021}
}

@article{Kebede2024,
  title={Assessing and addressing the global state of food production data scarcity},
  author={Kebede, Endalkachew Abebe and Abou Ali, Hanan and Clavelle, Tyler and Froehlich, Halley E and Gephart, Jessica A and Hartman, Sarah and Herrero, Mario and Kerner, Hannah and Mehta, Piyush and Nakalembe, Catherine and others},
  journal={Nature Reviews Earth \& Environment},
  pages={1--17},
  year={2024},
  publisher={Nature Publishing Group UK London}
}

@article{Kempf2014,
  title={New Ways of Mapping Knowledge Organization Systems: Using a Semi-Automatic Matching Procedure for Building up Vocabulary Crosswalks.},
  author={Kempf, Andreas Oskar and Ritze, Dominique and Eckert, Kai and Zapilko, Benjamin},
  journal={Knowledge Organization},
  volume={41},
  number={1},
  year={2014}
}

@misc{LD4D2023,
  author       = {{Livestock Data for Decisions}},
  title        = {Finding common ground: building shared solutions for livestock feed data},
  howpublished = {\url{https://livestockdata.org/news/finding-common-ground-building-shared-solutions-livestock-feed-data}},
  year         = {2023},
  month        = sep,
  note         = {Accessed: 2026-05-20}
}

@manual{Loria2014,
  title  = {{TextBlob}: Simplified Text Processing},
  author = {Loria, Steven},
  year   = {2014},
  note   = {Python library version 0.16.0},
  url    = {https://github.com/sloria/TextBlob}
}

@article{Raymond2023,
  title={Informatics progress of the Global Burden of Animal Diseases programme towards data for One Health.},
  author={Raymond, K and BenSassi, N and Patterson, GT and Huntington, B and Rushton, J and Stacey, DA and Bernardo, TM},
  journal={Revue Scientifique et Technique (International Office of Epizootics)},
  volume={42},
  pages={218--229},
  year={2023}
}

@article{Schrobback2023,
  title={Approximating the global economic (market) value of farmed animals},
  author={Schrobback, Peggy and Dennis, Gabriel and Li, Yin and Mayberry, Dianne and Shaw, Alexandra and Knight-Jones, Theodore and Marsh, Thomas Lloyd and Pendell, Dustin L and Torgerson, Paul R and Gilbert, William and others},
  journal={Global food security},
  volume={39},
  pages={100722},
  year={2023},
  publisher={Elsevier}
}

@misc{StatCan2025,
  author       = {{Statistics Canada}},
  title        = {32-26-0008},
  year         = {2025},
  howpublished = {\url{https://www150.statcan.gc.ca/n1/pub/32-26-0008/322600082025001-eng.htm}},
  note         = {Accessed September 4, 2026}
}

@article{Subirats2022,
  title={{AGROVOC} The linked data concept hub for food and agriculture},
  author={Subirats-Coll, Imma and Kolshus, Kristin and Turbati, Andrea and Stellato, Armando and Mietzsch, Esther and Martini, Daniel and Zeng, Marcia},
  journal={Computers and Electronics in Agriculture},
  volume={196},
  pages={105965},
  year={2022},
  publisher={Elsevier}
}

@article{Mulchandani2023,
  title={Global trends in antimicrobial use in food-producing animals: 2020 to 2030},
  author={Mulchandani, Ranya and Wang, Yu and Gilbert, Marius and Van Boeckel, Thomas P},
  journal={PLOS Global Public Health},
  volume={3},
  number={2},
  pages={e0001305},
  year={2023},
  publisher={Public Library of Science San Francisco, CA USA}
}

@misc{FAO_AGROVOC_2026,
  author       = {{Food and Agriculture Organization of the United Nations}},
  title        = {About | {AGROVOC}},
  howpublished = {\url{https://www.fao.org/agrovoc/about}},
  year         = {2026},
  note         = {Accessed: May 13, 2026}
}

@misc{USDA_NALT_2022,
  author       = {{U.S. Department of Agriculture, National Agricultural Library}},
  title        = {{NAL} Agricultural Thesaurus {(NALT)} Concept Space},
  howpublished = {\url{https://lod.nal.usda.gov/nalt/en/}},
  year         = {2022},
  note         = {Accessed: May 13, 2026. Last modified: July 16, 2024}
}

@techreport{Tripartate2022,
  author      = {{World Organisation for Animal Health} and {Food and Agriculture Organization of the United Nations} and {World Health Organization}},
  title       = {Surveillance and Information Sharing Operational Tool: An Operational Tool of the Tripartite Zoonoses Guide},
  institution = {WOAH, FAO, and WHO},
  year        = {2022},
  month       = sep,
  address     = {Geneva, Rome, and Paris},
  isbn        = {978-92-4-005325-0},
  url         = {https://www.woah.org/app/uploads/2022/09/surveillance-and-information-sharing-operational-tool-eng.pdf}
}

@article{Wilkinson2016,
   author = {Mark D. Wilkinson and Michel Dumontier and IJsbrand Jan Aalbersberg and Gabrielle Appleton and Myles Axton and Arie Baak and Niklas Blomberg and Jan Willem Boiten and Luiz Bonino da Silva Santos and Philip E. Bourne and Jildau Bouwman and Anthony J. Brookes and Tim Clark and Mercè Crosas and Ingrid Dillo and Olivier Dumon and Scott Edmunds and Chris T. Evelo and Richard Finkers and Alejandra Gonzalez-Beltran and Alasdair J.G. Gray and Paul Groth and Carole Goble and Jeffrey S. Grethe and Jaap Heringa and Peter A.C. t Hoen and Rob Hooft and Tobias Kuhn and Ruben Kok and Joost Kok and Scott J. Lusher and Maryann E. Martone and Albert Mons and Abel L. Packer and Bengt Persson and Philippe Rocca-Serra and Marco Roos and Rene van Schaik and Susanna Assunta Sansone and Erik Schultes and Thierry Sengstag and Ted Slater and George Strawn and Morris A. Swertz and Mark Thompson and Johan Van Der Lei and Erik Van Mulligen and Jan Velterop and Andra Waagmeester and Peter Wittenburg and Katherine Wolstencroft and Jun Zhao and Barend Mons},
   doi = {10.1038/sdata.2016.18},
   issn = {2052-4463},
   issue = {1},
   journal = {Scientific Data 2016 3:1},
   month = {3},
   pages = {1-9},
   pmid = {26978244},
   publisher = {Nature Publishing Group},
   title = {The {FAIR} Guiding Principles for scientific data management and stewardship},
   volume = {3},
   url = {https://www.nature.com/articles/sdata201618},
   year = {2016}
}

@article{Zeng2008,
  title={Knowledge organization systems {(KOS)}},
  author={Zeng, Marcia Lei},
  journal={KO Knowledge Organization},
  volume={35},
  number={2-3},
  pages={160--182},
  year={2008},
  publisher={Nomos Verlagsgesellschaft mbH \& Co. KG}
}
\end{document}